\documentclass[conference]{IEEEtran}
\usepackage[utf8]{inputenc}
\IEEEoverridecommandlockouts
\usepackage{cite}
\usepackage{amsmath,amssymb,amsfonts}
\usepackage{algorithmic}
\usepackage{graphicx}
\usepackage{textcomp}
\usepackage{xcolor}
\usepackage{newunicodechar}
\newunicodechar{−}{\ensuremath{-}}
\def\BibTeX{{\rm B\kern-.05em{\sc i\kern-.025em b}\kern-.08em
    T\kern-.1667em\lower.7ex\hbox{E}\kern-.125emX}}
\begin{document}

\title{A Quantum Reservoir for
\\Neurodynamical Forecasting}

\author{\IEEEauthorblockN{Annemarie Wolff}
\IEEEauthorblockA{\textit{Uni. de Montr\'eal} \\
\textit{Centre Hospitalier Univ. de Sainte-Justine}\\
Montr\'eal, PQ, CA \\
anne-marie.wolff@umontreal.ca}
\and
\IEEEauthorblockN{ Kathleen Hamilton}
\IEEEauthorblockA{\textit{Oak Ridge National Laboratory} \\
Oak Ridge, TN, USA \\
hamiltonke@ornl.gov}
\and
\IEEEauthorblockN{Kahn Rhrissorrakrai}
\IEEEauthorblockA{\textit{IBM Quantum} \\
Yorktown Heights, NY, USA \\
krhriss@us.ibm.com
}
\and
\IEEEauthorblockN{Laxmi Parida}
\IEEEauthorblockA{
\textit{IBM Quantum}\\
Yorktown Heights, NY, USA \\
parida@us.ibm.com}
\and
\IEEEauthorblockN{Filippo Utro}
\IEEEauthorblockA{\textit{IBM Quantum} \\
Yorktown Heights, NY, USA \\
futro@us.ibm.com}
\and
\IEEEauthorblockN{Guillaume Dumas}
\IEEEauthorblockA{\textit{Uni. de Montr\'eal} \\
\textit{Centre Hospitalier Univ. de Sainte-Justine}\\
\textit{Mila – Quebec AI Institute}\\
Montr\'eal, PQ, CA \\
guillaume.dumas@umontreal.ca}
}

\maketitle

\begin{abstract}
Forecasting neural activity from short recordings remains a fundamental challenge. Reservoir computing may offer an efficient paradigm for temporal prediction, however classical reservoirs typically underperform in small‑data regimes. Here we investigate whether quantum reservoir computing (QRC) can help overcome this limitation. Building on recent advances, we introduce a quantum reservoir based on a transverse‑field Ising model, combined with heterogeneous quantum measurements and polynomial ridge regression. On a standard benchmark task, results show that the quantum reservoir outperforms a classical counterpart overall, with prediction accuracy strongly dependent on reservoir parameters. We further demonstrate feasibility by running the same task on quantum hardware. To assess performance on biological signals, we evaluate QRC on simulated human electroencephalography (EEG) data with a parallel reservoir architecture. On this challenging task, the tested quantum reservoir did not match the performance of the classical one, but it produced stable, convergent predictions. This is a meaningful first step toward forecasting of biologically realistic neural data using a quantum reservoir. Overall, our findings indicate that although current quantum hardware and parallel reservoir architectures do not yet surpass classical methods on complex neural signals, QRC can be executed on near‑term devices and does converge with realistic EEG‑like data. This work establishes a practical baseline for future algorithmic and hardware developments aimed at clinical time‑series forecasting with quantum systems.
\end{abstract}


\section{Introduction}
Understanding and predicting complex neural activity remains a significant challenge in modern neuroscience ~\cite{Breakspear2017BrainDynamics, ramezanian-panahi2022generative}. This has several implications. Accurate prediction of an individuals neural activity is gaining importance in both basic neuroscience and clinical applications~\cite{Shenoy2021BrainWide}. However, understanding the structure of the brain and cortical areas alone is insufficient for this task; even a complete map of a neural circuit does not allow reliable prediction of its resulting  activity~\cite{Beiran2025ConnectomeRNN}. As a result, the dynamics must be learned directly from the neural data itself, but the ability to do so is currently hampered by the limited volume of available data~\cite{Vlachas2020ForecastingRC}.

One approach that works well for complex dynamical systems is reservoir computing (RC). RC comprises a randomly connected fixed recurrent neural network (the reservoir) where only the linear readout layer is trained~\cite{Jaeger2004Harnessing, Maass2002LiquidState}, yet it can approximate almost any nonlinear system with fading memory~\cite{Lukosevicius2009RCReview}. Although these characteristics pair well with the analysis of neural data, the low volume of data for training, and the slow long-range temporal correlations present\cite{Honey2012SlowCortical} still pose a significant challenge for RC.

Quantum reservoir computing (QRC) leverages quantum-mechanical properties for computation \cite{Fujii2017DisorderedQRC} and offers several potential advantages that are particularly relevant for neural modeling under data scarcity. Classical reservoirs typically require large training sets to achieve good generalization, but quantum reservoirs can learn effectively from fewer training samples \cite{Ahmed2025RobustQRC}. Quantum dynamics generate rich feature spaces with better sample efficiency; a small number of qubits can generate very high‑dimensional features, while a classical reservoir of comparable size would be severely limited in representational capacity \cite{GarciaBeni2023PhotonicQRC}. They also exhibit slower memory decay due to non‑Markovian effects \cite{Sannia2025NonMarkovianQRC}. Together, these properties suggest that QRC may be particularly well‑suited for the small‑data, long‑memory regime that characterizes much of neural time‑series forecasting.

\begin{figure}
    \centering
    \includegraphics[width=\linewidth]{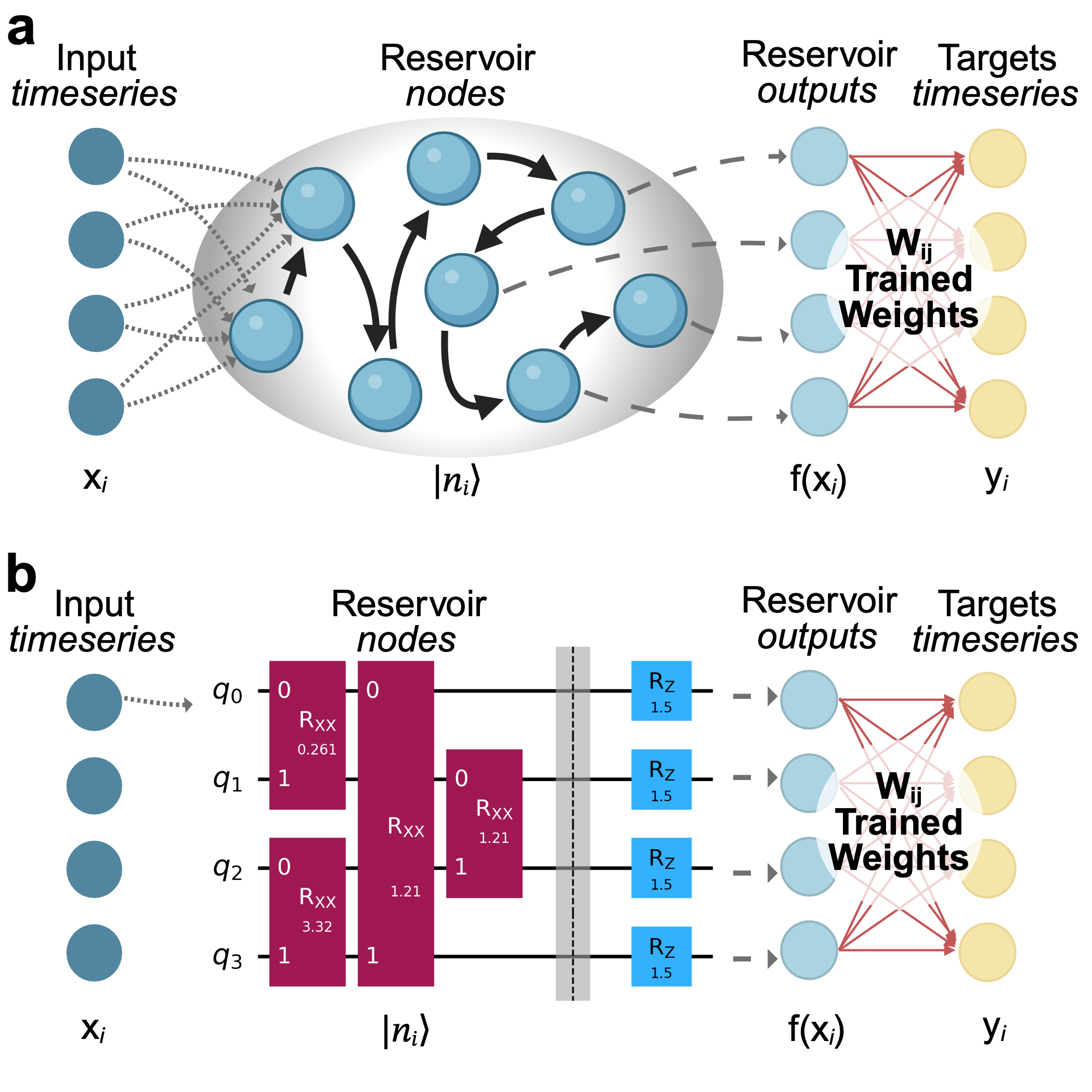}
    \caption{Quantum reservoir with trained ridge regression. (a) The time series data $x_{i}$ (left) was input to the reservoir (gray dotted arrows). The nodes in the reservoir are coupled according to $k$, a tuned parameter. The system evolves for time $dt$, and the state of the nodes becomes the outputs (gray dashed arrows). Ridge regression calculates the trained weights $W_{ij}$ from the reservoir outputs $f(x_{i})$ and the target time series $y_{i}$. (b) In the quantum implementation, the quantum circuit serves as the reservoir, with the time series input encoded into the circuit on qubit $q_{0}$, and the measurement of the states occurring on all qubits.}\label{fig:overview}
\end{figure}

To examine this directly, we test a quantum reservoir to predict neural activity (Fig.~\ref{fig:overview}). We implemented a quantum reservoir and compared its performance with a classical counterpart, an Echo State Network (ESN).
While the two architectures differ fundamentally in state representation and feature generation, we enforced comparability by using standard parameter ranges for ESN's, fixing all data preprocessing, training, and evaluation, and by optimizing hyperparameters separately for each model using the same multi-objective criterion.

Our goal was to determine whether the theoretical benefits of QRC are evident for the prediction of neural time series data, and to identify when quantum approaches are beneficial. We first tested performance on a common reservoir prediction benchmark, a superimposed oscillator time series. We then tested a biologically realistic simulation of human EEG data with a parallel reservoir architecture.

\section{Methods}

\subsection{Signal generation and simulation}
All analyses were done with Python (v3.10.13). Simulated signals were generated using a multiple superimposed oscillator model consisting of two sinusoidal components at 1.0Hz and 2.5Hz, each with unit amplitude. To each signal simulation (25 repetitions), we added zero-mean Gaussian noise with a standard deviation of 0.05 and a unique random seed. We then simulated a biologically realistic human electroencephalography (EEG) signal, which contained two oscillations (6Hz and 10Hz, jitter of 0.5Hz). These signals also contained zero-mean Gaussian noise (SD = 0.1) and 1/$f$ (pink) noise (amplitude = 0.5). All signals were simulated at 50Hz for 60 seconds.

\subsection{Data preprocessing and normalization}
Both reservoirs used identical preprocessing to ensure a fair comparison. Reservoir inputs were normalized in two stages. First, the data was z-scored using the training-data mean and standard deviation. Second, an $\arctan$ transform was applied to softly bound the z-scored values to [−1, 1]. The data was then shifted and rescaled to [0, 1] for the reservoir input encoding.

Normalization parameters were computed solely from the training data and held fixed for testing. In contrast, the regression targets were kept on the z-scored scale, without the $\arctan$ step. This was done so that predictions were recovered with a single linear rescaling back to the original mean and standard deviation. Keeping the targets linear avoids the numerical instability that an inverse-tangent recovery can produce near the boundaries of the [0, 1] interval.

\subsection{Quantum reservoir}
In reservoir computing, a sufficiently complex, nonlinear dynamical system can project high-dimensional input signals into a feature space where simple regression can extract the desired output (Fig.~\ref{fig:overview}). The quantum implementation of a reservoir uses the principles of quantum mechanics to generate the underlying dynamical system and generate features. Following previous implementations of quantum reservoirs for time series prediction~\cite{Fujii2017DisorderedQRC, Kutvonen2020OptimizingQRC}, we employed a quantum reservoir that evolves according to a transverse-field Ising model. This model is a canonical quantum many-body system in which qubits (spins) interact with their neighbors while being subject to a perpendicular magnetic field. It generates complex, tunable dynamics that are well-suited for reservoir computing and is described by a Hamiltonian $H$,
\begin{equation}
    H = \sum_{i,j}J_{i,j}X_iX_j+\sum_{i}h_iZ_i
\end{equation}
where an interaction term represents spin-spin interactions along the transverse direction, and a single-qubit term applies a magnetic field perpendicular to the natural quantization axis. The interaction term has coupling strengths $J_{ij}$ between coupled qubits $i$ and $j$ and applies an X-basis coupling. As described in Section II-D, only neighboring qubits are coupled.

The second field term is applied along the Z-basis on each qubit, and describes coupling to an external magnetic field of strength $h_i$ along the longitudinal axis. The coupling between the spins was set to
\begin{equation}
    J_{i,j}^k=\frac{(i+j)^k}{c_k}
\end{equation}
where $i$ and $j$ are the spin indices, $k$ is a scaling parameter, and
\begin{equation}
    c_k = \frac{\frac{2}{N_s(N_s-1)} \sum_{i<j} (i+j)^k}{0.5}
\end{equation}
where $c_k$ is a $k$-dependent constant which ensures comparable energy scales between setups, with the mean interaction $E[J^k] = 0.5$. Because couplings are placed only between neighboring qubits (see Section II-D), the exponent $k$ does not set a decay with distance. Instead, with the mean coupling held fixed at 0.5 by the normalization, $k$ controls how unevenly the coupling strength is distributed across the ring's neighbor links, from nearly uniform at small $k$ to strongly graded at large $k$. This heterogeneity shapes how information propagates and mixes through the reservoir, and is the property we tune.

We varied $k \in \{1,2,3,4\}$ for the superimposed oscillators. The reservoir size was also varied, using 4 and 5 qubits (one injection qubit plus three or four memory qubits). Together with the four coupling exponents $k$ and five evolution times $dt$, this gave 40 quantum reservoir configurations.

As varying magnetic field strength had no effect on prediction accuracy~\cite{Kutvonen2020OptimizingQRC}, magnetic field strength $h$ was set uniformly to 0.5 on all qubits.

\subsection{Quantum circuit construction}
To minimize circuit depth on quantum hardware, we used a ring topology instead of all-to-all coupling. Specifically, we partitioned the coupling gates into layers such that gates within each layer act on non-adjacent pairs of qubits. As a result, the reservoir evolution can be implemented with two or three layers of parallelized unitaries.

In addition, we adopted a rewind protocol~\cite{hamhoum2025multivariate} that accounts for the noise in deeper circuits on current superconducting hardware. At each timestep $t$, an input value drives the quantum state evolution. For a density matrix $\rho(t)$ representing the quantum state, evolution over time interval $\Delta t$ is given by
\begin{equation}
    \rho(t+\Delta t)=e^{-i\Delta tH}\rho(t)e^{i\Delta tH}
\end{equation}
The evolution is carried out in one timestep~\cite{Kutvonen2020OptimizingQRC}, with the full Hamiltonian applied. We used one Trotter step~\cite{Kutvonen2020OptimizingQRC} to decrease the depth of our circuit. The evolution time $dt$ (arbitrary units) controls the overall evolution time per input, and we varied it with the following steps for superimposed oscillations, $dt \in \{0.1,0.5,1.0,3.0,5.0\}$.

Our simulated data was applied to two circuit constructions. The first was one reservoir for single frequencies and the superimposed oscillators task. In this reservoir, the $dt$ and $k$ values were consistent across the ring, which used the same nearest neighbor $RXX$ connectivity as the parallel construction. The second construction, for the biologically realistic multi-frequency simulated EEG data, was with three small reservoirs that were independent. The reservoir outputs were combined to calculate the weights for ridge regression. This was done as there were two frequencies in the signal ($\theta$ = 6 Hz, $\alpha$ = 10 Hz) and 1/$f$ noise, so each signal component had its own reservoir with its own particular $dt$ and $k$ values~\cite{manneschi2021parallelESN}.

\subsection{Input encoding and state readout}
The $n$-qubit register is prepared in the state $|+\rangle^{\otimes n}$. 
The input time series $s_i$ is normalized 
and injected into the reservoir 
at time $t_i$ by preparing the first qubit in the state 
\begin{equation}
    |\psi_{s_i}\rangle=\sqrt{(1-s_i)}|0\rangle+\sqrt{s_i}|1\rangle
\end{equation}
The system evolves under Hamiltonian $H$ for a duration $\Delta t$ until the next input arrives. During this interval, information encoded in qubit$_{0}$ spreads through the network via the inter-qubit couplings, creating input-dependent signatures throughout the many-body state.

We measured all three Pauli operators (X, Y, and Z) on each qubit and computed $\langle X\rangle, \langle Y\rangle, \langle Z\rangle$~\cite{Zhu2025FewAtomQRC}. Computing these three measurements captures information about both the diagonal elements (revealed by the Z measurement) and the off-diagonal elements (capturing coherence in the quantum state, as revealed by the X and Y measurements). This multi-basis measurement approach~\cite{Zhu2025FewAtomQRC} enriches the feature space available to the readout layer without requiring additional qubits. During the evolution interval [$t_i$, $t_i + \Delta t$], we measure $\langle Z_i \rangle$ on the readout nodes. The expectation values are estimated using the \texttt{Estimator} primitive in Qiskit (v2.1.1) \cite{javadiabhari2024quantumcomputingqiskit}.

To extract high-dimensional feature vectors from a modest number of physical qubits with the simulator, we employ temporal multiplexing~\cite{Nakajima2014SoftBodyRC} in which we divide $\Delta t$ into $N_v$ equal time steps, effectively creating virtual nodes~\cite{Larger2017PhotonicRC} from a single physical network. 

The raw readout features are the vector of estimated expectation values that is passed to the readout layer, augmented with polynomial cross-products~\cite{Zhu2025FewAtomQRC}. This quadratic augmentation creates new features from the raw measurements by taking all pairwise products of the original measurement values. The reason is that the quantum reservoir's dynamics may not linearly separate the desired output. Adding these quadratic terms gives the subsequent ridge regression access to nonlinear combinations of the reservoir state, improving prediction \cite{Zhu2025FewAtomQRC}.

Finally, in the simulation implementation (Qiskit Aer v0.17.1) of the superimposed oscillator task, the noise model was used for the corresponding hardware IBM Heron R2 ($ibm\_quebec$). For each repetition of the parameter combination (25 repetitions), a set of connected qubits was randomly chosen. The noise profiles from these qubits were applied to each evolution of the quantum circuit for that repetition. This was done to find the best parameter combinations in the presence of realistic hardware noise while varying over repetitions.

\subsection{Circuit transpilation and physical qubit selection}
We employed stochastic transpilation prior to analysis to minimize the circuit depth executed on hardware. A reference reservoir circuit was constructed using $n$ qubits and
all input encoding angles set to a neutral reference value (input = 0.5, $\theta = \pi/2$). This reference circuit was transpiled 100 times (optimization level 3), with a unique integer seed (0–99) for each repetition. Pauli Twirling was applied to all entangling gates after transpilation. This is a noise mitigation technique that randomizes coherent errors into simpler stochastic noise. The lowest two-qubit gate depth was then selected and used directly for the analysis.
The inclusion of stochastic transpilation and Pauli twirling in the quantum implementation are noise-mitigation steps that can improve the fidelity of quantum circuit executions on hardware. 
These methods do not influence predictive accuracy.

\subsection{Classical Echo State Network}
An Echo State Network (ESN) is a type of recurrent neural network. It is composed of a fixed, randomly initialized recurrent layer (reservoir) whose weights are never modified during training, and a single linear output layer (readout) whose weights are trained via linear regression. As with the quantum reservoir, the classical readout used ridge regression with identical polynomial (quadratic) feature augmentation \cite{Zhu2025FewAtomQRC}, ensuring the two architectures were compared under the same readout.

Just as in the quantum implementation, two hyperparameters were varied in the classical ESN, the spectral radius and the leak rate. The leak rate controls temporal memory (lower leak rate = slower dynamics, longer memory), and the spectral radius governs stability and dynamical richness (values near 1 maximize complexity without instability). Although no direct mathematical mapping exists to quantum parameters, functionally evolution time $dt$ parallels the leak rate (both control temporal integration), and coupling exponent $k$ parallels the spectral radius (both shape the richness of the reservoir dynamics). The spectral radius was varied from 0.4 to 2.0 (step = 0.4), and the leak rate was varied from 0.2 to 1.0 (step = 0.2). The input scaling parameter was fixed at 1.0. 

\subsection{Reservoir warm-up}
Both the quantum and classical reservoirs employed an identical two-pass warm-up strategy, originally developed for ESNs~\cite{Lukosevicius2012PracticalESN}. In the first pass, the reservoir was driven through the first 200 training inputs without collecting any readout values. Once the warm-up phase was complete, and without re-initializing the reservoir state, the system was run from the beginning of the training sequence, collecting readout measurements at every timestep.

\subsection{Ridge regression and autoregressive prediction}
Reservoir features were organized into a matrix where each row contains the feature vector $\hat{Y} = [\hat{y}(t)]$
and target values (the one-step-ahead training pairs) were similarly organized as a vector. 
Ridge regression, with a bias term and polynomial feature augmentation, was used to predict the targets from the features~\cite{Zhu2025FewAtomQRC}. During testing, predictions were generated autoregressively using the trained readout weights. 
 Starting from the final training sample, the reservoir was run forward one timestep at a time, with the previously predicted value as the input. At each step, the reservoir maintained its continuously evolving internal state. The readout weights were applied to compute a prediction in normalized space and this prediction was inverse-normalized to recover a value on the original signal scale. This closed-loop prediction tested whether the reservoir could sustain coherent dynamics over an extended horizon when driven by its own predictions.

\subsection{Performance metrics}
Both systems were evaluated using three metrics. The first is Normalized Mean Squared Error (NMSE)~\cite{Jaeger2001EchoState, Lukosevicius2012PracticalESN}
\begin{equation}
    \mathrm{NMSE} = \frac{\dfrac{1}{N} \sum_t \bigl [y(t) - \hat{y}(t)\bigr ]^2}{\mathrm{\sigma^2}(y)}
\end{equation}
where $y(t)$ is the target signal, $\hat{y}(t)$ is the reservoir prediction, $N$ is the number of samples and $\mathrm{\sigma^2}(y)$ is the variance of the target signal.

The second metric calculated was dynamic time warping (DTW)~\cite{Sakoe1978DTW}. It finds the optimal non-linear alignment between two sequences by minimizing the cumulative Euclidean distance across all possible monotonic warping paths.

The final metric computed was phase-locking value (PLV)~\cite{Lachaux1999PhaseSynchrony}. Each signal was first bandpass-filtered using a 4th-order zero-phase Butterworth filter spanning $\frac{1}{2}$  the minimum to 2 times the maximum signal frequency, isolating the frequency content of interest. The instantaneous phase, $\phi$, was then extracted via the Hilbert transform. PLV was computed as
\begin{equation}
    \mathrm{PLV}= \frac{1}{N}\left| \sum_{t=1}^N e^{i(\phi(t)-\hat{\phi}(t))}\right|
\end{equation}
where $\phi$ and $\hat{\phi}$ are the instantaneous phase time series of the target and predicted signals, respectively, and $N$ is the number of samples. To maintain a consistent directionality across all metrics (0 equates to no error), PLV was converted to 1-PLV.

Repetitions with NMSE $\geq$ 1 (no better than predicting the mean of the signal) were considered non-converging and excluded from the aggregate comparison for both reservoirs.


\subsection{Identification of optimal parameter combinations}
To determine the optimal combination of accuracy metrics to be run on the quantum hardware, and for direct comparison with the classical reservoir, we computed the Pareto frontier. Using three variables (DTW, NMSE, 1-PLV), the median (robust to outliers) and standard deviation of each over all repetitions (6 objectives) were used to calculate the multi-objective Pareto-dominance ranking. Standard deviation was included as we wanted to select the hyperparameters based not only on accuracy, but also on consistency. The Pareto-optimal frontier (rank 1) comprised all combinations not dominated by any other combination. Non-dominated sorting was performed using the non-dominated sorting algorithm of NSGA-II~\cite{Deb2002NSGAII}, implemented via the \textit{pymoo} toolbox~\cite{Blank2020Pymoo}.


\section{Results}
\subsection{Super-imposed oscillators}
\begin{figure*}[!t]
    \centering
    \includegraphics[width=1\textwidth]{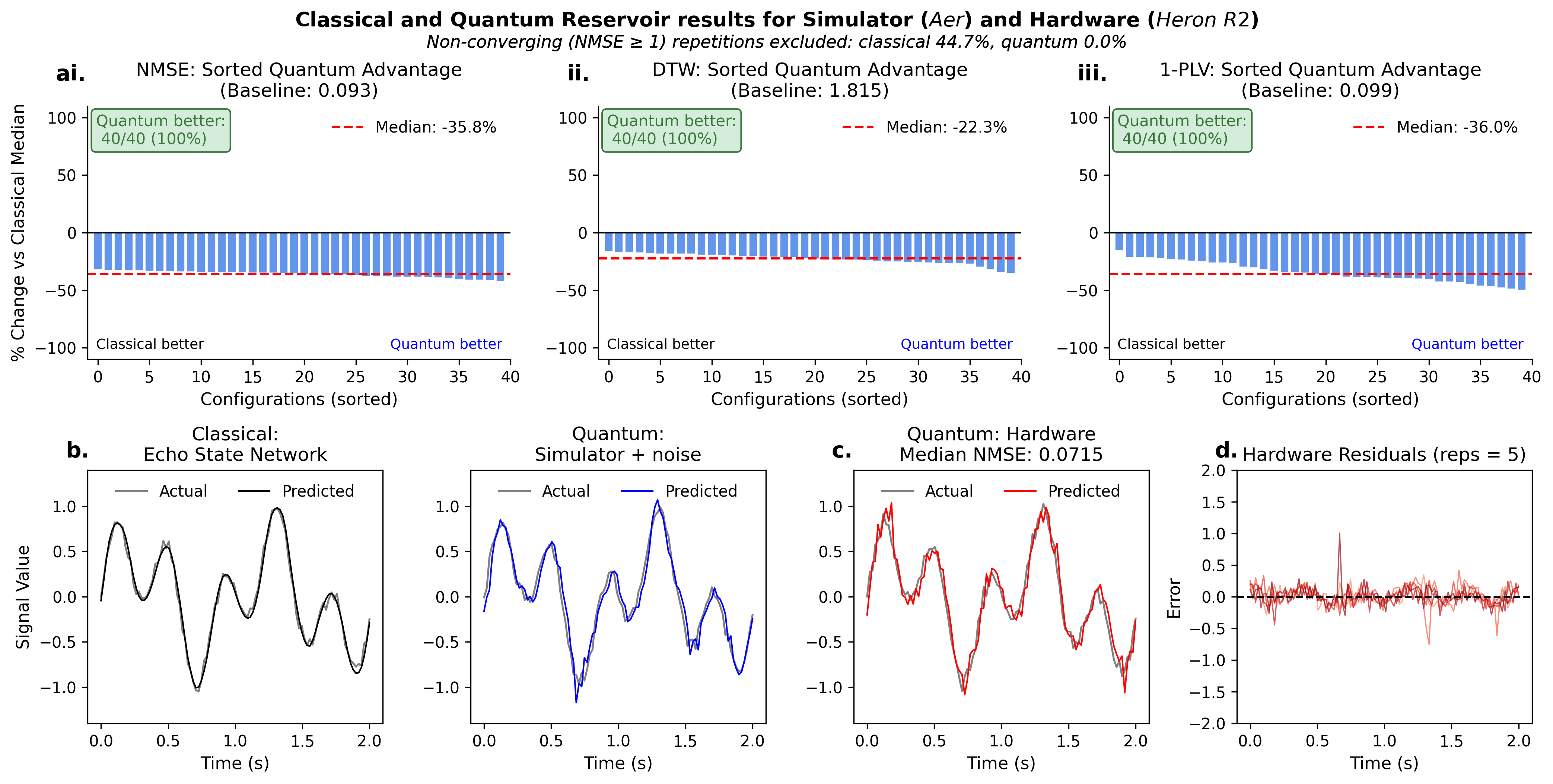}
    \caption{Benchmark task and simulated biological data results.  We compared the classical and quantum reservoir results (median) across all parameter combinations. (a.i.) The normalized mean squared error (NMSE) metric showed better results (lower NMSE) in the quantum simulator implementation (blue) in 40 (100\%) of the parameter combinations. (a.ii.) Dynamic time warping (DTW) was found to have better results in the quantum implementation in 40 (100\%) of the parameter combinations. (a.iii.) Phase-locking value (PLV), which measures the phase consistency of the signal, was found to be better in the quantum implementation in 40 (100\%) of the parameter combinations. Classical (b, left) and quantum simulator (b, right) results for one repetition and set of parameters. The predicted time series (classical = black, quantum = blue) fit the actual time series (gray) in both regimes for the superimposed oscillators benchmark task. (c) When the quantum reservoir was tested on the IBM Heron R2 hardware, the results were consistent (repetition with median NMSE shown). (d) Over 5 repetitions, the residual errors clustered around zero, with some variability at the peaks and troughs. }
    \label{fig:results}
\end{figure*}

On the superimposed oscillators benchmark task (Fig.~\ref{fig:results}a) we found that overall the quantum simulator implementation performed better than the classical implementation. We compared the median performance (over 25 repetitions) of each of 40 quantum reservoir configurations (2 qubit counts x 4 coupling exponents $k$ x 5 evolution times $dt$) against the pooled median of the classical reservoir (2 node counts x 4 spectral radius levels x 5 leak rate levels). The median error in the quantum implementation on the simulator was lower (better) than the classical baseline for all 40 configurations on NMSE, DTW, and 1-PLV, with a median improvement of 35.8\%, 22.3\%, and 36.0\% respectively.

The quantum reservoir was also more robust across the 25 repetitions. No quantum repetition failed to converge (repetitions with NMSE $\geq$ 1), whereas 44.7\% of classical repetitions did. Among converging runs, the quantum reservoir showed lower per-configuration variability on all three metrics (median NMSE SD 0.006 vs. 0.136; DTW 0.14 vs. 0.74; 1$-$PLV 0.035 vs. 0.096).


We next ran the time series prediction on the quantum hardware, with several necessary changes. Our analysis was encoded onto the hardware in a sliding window of 9 datapoints (no overlap). There were 2 evolution steps per window, with no intermediate readouts, and a single measurement (5,000 shots) at the end of each window. Pauli Twirling (100 randomizations) was applied for noise mitigation, and the transpilation optimization level was 3. Five repetitions of the time series prediction, with the optimal parameter combination, were run on the hardware (Fig.~\ref{fig:results}c,d). The median results were an NMSE of 0.0715, a DTW of 1.8481, and a 1-PLV of 0.0472.

\begin{table}[t]
\caption{Best parameters for Classical implementation \\of superimposed oscillators benchmark task}
\label{tab:best_configs}
\centering
\scriptsize
\setlength{\tabcolsep}{3pt}
\begin{tabular}{cccccccc}
\hline
Rank & Nodes*& Spectral Radius& Leak Rate& Windows& NMSE & DTW & 1--PLV \\
\hline
1 & 250 & 2.0 & 0.7 &  5 & 0.0380 & 1.1305 & 0.0256 \\
2 & 250 & 2.0 & 0.7 & 10 & 0.0180 & 1.0461 & 0.0394 \\
3 & 250 & 1.2 & 0.7 & 15 & 0.0116 & 0.9305 & 0.0266 \\
4 & 250 & 2.0 & 0.5 & 15 & 0.0187 & 0.9903 & 0.0372 \\
5 & 250 & 2.0 & 0.7 & 15 & 0.0169 & 1.0044 & 0.0992 \\
\hline
\end{tabular}
    \vspace{0.1cm}
  \begin{minipage}{0.95\linewidth}
    \textit{*Best configuration shown; 250 and 375 nodes were tested}.
  \end{minipage}
\end{table}

\begin{table}[t]
\caption{Best parameters for Quantum implementation \\of superimposed oscillators benchmark task
}
\label{tab:qrc_best}
\centering
\scriptsize
\setlength{\tabcolsep}{3pt}
\begin{tabular}{cccccccc}
\hline
Rank & Qubits*& Evolution ($dt$)& Coupling ($k$)& Windows& NMSE & DTW & 1--PLV \\
\hline
1 & 4 & 0.1 & 3 & 10 & 0.0618 & 1.3503 & 0.0549 \\
2 & 4 & 0.1 & 4 & 10 & 0.0602 & 1.2832 & 0.0489 \\
3 & 4 & 0.5 & 3 & 10 & 0.0610 & 1.4428 & 0.0423 \\
4 & 4 & 0.5 & 4 & 10 & 0.0587 & 1.4116 & 0.0488 \\
5 & 4 & 1.0 & 2 & 10 & 0.0556 & 1.3884 & 0.0446 \\
\hline
\end{tabular}
\vspace{0.1cm}
  \begin{minipage}{0.98\linewidth}
    \textit{*Best configuration shown; 4 and 5 qubits were tested}.
  \end{minipage}
\end{table}

\subsection{Biologically realistic multi-frequency EEG simulation}

For the simulated EEG data, we did a preliminary analysis of simulated sinusoids (6 Hz, 10 Hz) to determine the best parameters. For each sub-reservoir, parameter combinations were selected from the corresponding single-frequency Pareto analysis (6 Hz for the $\theta$ reservoir; 10 Hz for the $\alpha$ reservoir; fixed configuration for the 1/$f$ reservoir). The number of training windows was consistent across both implementations, from 5 to 15 windows (step = 1).

After Pareto frontier analysis, the best parameter combination was $\mathrm{spectral\ radius}_\alpha$ of 0.8, $\mathrm{leak\ rate}_\alpha$ of 1.0, $\mathrm{spectral\ radius}_\theta$ of 1.2, $\mathrm{leak\ rate}_\theta$ of 0.8, $\mathrm{spectral\ radius}_f$ of 1.2, $\mathrm{leak\ rate}_f$ of 0.6, and 10 training windows. 

\begin{table}[t]
\caption{Multi-Frequency simulated EEG Results (simulator)}
\label{tab:eeg}
\centering
\begin{tabular}{lcccc}
\hline
Implementation & Windows & NMSE*& DTW*& 1--PLV*\\
\hline
Classical & 10& 0.0330& 1.787& 0.0110\\
Quantum   & 9& 0.4208& 5.231& 0.1880\\
\hline
\end{tabular}
\vspace{0.1cm}
  \begin{minipage}{0.85\linewidth}
    \textit{*Median over repetitions.}
  \end{minipage}
\end{table}

For the quantum simulator implementation, each of the smaller reservoirs had 4 qubits (3 $\times$ 4 = 12 qubits). For the smaller $\alpha$ reservoir, the $dt_{\alpha}$ was 2 and the $k_\alpha$ was 1. There were two sets of parameters to test for the smaller $\theta$ reservoir, specifically the $dt_{\theta}$ were 1 and 3 and the $k_\theta$ were 1 and 2. For the smaller reservoir for the 1/$f$ noise was a $dt_f$ of 5 and $k_f$ of 1.

The Pareto frontier analysis identified the best parameter combinations. They were a $k$-value of 1 for all smaller reservoirs, a $dt_{\alpha} = 2$, a $dt_{\theta}=1$, and a $dt_{pink}=5$. The number of training windows was 9. Finally, to compare the classical and quantum analysis of the multi-frequency EEG signal, a MANOVA found a significant difference in the dependent variables between the two implementations (F(3,46) = 474.25, $p \leq 0.001$, Pillai’s V = 0.969, partial $\eta^{2}$ = 0.97), with the classical reservoir outperforming the quantum implementation on this task.

\section{Discussion}
This study evaluated a transverse-field Ising quantum reservoir on two autoregressive forecasting tasks and demonstrated execution of the forecasting pipeline on superconducting quantum hardware. The results reveal two distinct regimes. On the superimposed-oscillator benchmark, quantum reservoirs performed better than classical ones and were more reliable, and converged with comparable performance on the quantum hardware. On the simulated EEG task, however, the classical reservoir performed better than the quantum implementation on the simulator. 
The observed advantage is primarily one of robustness rather than accuracy. The fixed unitary dynamics of the transverse-field Ising reservoir generate a high-dimensional, well-conditioned feature set in a $2^n$-dimensional Hilbert space, read out across three measurement bases (X, Y, Z) to capture both populations and coherences.

Despite this robustness, accuracy on the simulated EEG signal was reduced compared to the super-imposed oscillators task. Because the classical implementation used the same parallel decomposition yet performed well, the possible cause may be the limited per-component capacity of the quantum sub-reservoirs. No single reservoir observed the full signal, so cross-frequency structure was recovered only at the linear readout. This may limit feature richness relative to the 250-node classical sub-reservoirs and the single-reservoir benchmark.

This is an early result from one type of reservoir architecture. A theoretical motivation for continued work is that an $n$-qubit quantum system evolves in a $2^{n}$-dimensional Hilbert space. While actual readout features are limited by accessible physical measurements (local observables), this exponential state space can still yield richer feature representations than a classical reservoir of comparable size. Whether this translates into a practical benefit which matters in clinical settings with short recordings, such as comparable accuracy from fewer training samples, is an open question. This work is a first step toward answering it.

Future work will therefore focus on quantum hardware implementation for EEG signals, circuit optimization, and possibly revisiting previous design choices~\cite{Kutvonen2020OptimizingQRC}. Extending this evaluation to recorded human EEG and to signals with richer, non-stationary dynamics than the simulated case studied here is also a priority. Beyond time-series forecasting in neural data, whether this theoretical benefit materializes in practice toward clinical applications in psychiatry and alternative mathematical frameworks for neural activity remains an open question for future work ~\cite{wolff2025quantum}.



\vspace{0.25cm}
\section*{Acknowledgments}

This work was partially funded by the Institute of Data Valorization, Montréal; the Canada First Research Excellence Fund (CF00137433); the FRQNT Strategic Clusters Program (Centre UNIQUE); FRQS (2024-2025–CB–350516); NSERC (DCECR-2023-00089); the Canadian Neurodevelopmental Research Training (CanNRT) Platform.
\vspace{0.15cm}

Xavier Hinaut (INRIA) and Ibrahim Shehzad (IBM Quantum) provided early pieces of advice for this paper, with help from Camille Brun-Jolicoeur (IBM Quantum). All computations were made possible by the Digital Research Alliance of Canada, Calcul Québec, and $PINQ^{2}$.
\vspace{0.15cm}

This manuscript has been partially authored by UT-Battelle, LLC, under Contract No. DE-AC0500OR22725 with the U.S. Department of Energy. The United States Government retains and the publisher, by accepting the article for publication, acknowledges that the United States Government retains a non-exclusive, paid-up, irrevocable, world-wide license to publish or reproduce the published form of this manuscript, or allow others to do so, for the United States Government purposes. The Department of Energy will provide public access to these results of federally sponsored research in accordance with the DOE Public Access Plan.
\vspace{0.15cm}

The authors acknowledge useful conversations and feedback within the Quantum Working Group for Health Care and Life Sciences\cite{bose_advancing_2026, flother_how_2024, doga_how_2024}.

\newpage

\bibliographystyle{plain}
\bibliography{references}

@misc{Ahmed2025RobustQRC,
  author       = {Ahmed, O. and Tennie, F. and Magri, L.},
  title        = {Robust Quantum Reservoir Computers for Forecasting Chaotic Dynamics: Generalized Synchronization and Stability},
  year         = {2025},
  eprint       = {2506.22335},
  archivePrefix= {arXiv},
  doi          = {10.48550/arXiv.2506.22335}
}

@article{Beiran2025ConnectomeRNN,
  author  = {Beiran, M. and Litwin-Kumar, A.},
  title   = {Prediction of Neural Activity in Connectome-Constrained Recurrent Networks},
  journal = {Nature Neuroscience},
  volume  = {28},
  number  = {1},
  pages   = {2561--2574},
  year    = {2025},
  doi     = {10.1038/s41593-025-02080-4}
}

@article{Blank2020Pymoo,
  author  = {Blank, J. and Deb, K.},
  title   = {Pymoo: Multi-Objective Optimization in Python},
  journal = {IEEE Access},
  volume  = {8},
  pages   = {89498--89509},
  year    = {2020},
  doi     = {10.1109/ACCESS.2020.2990567}
}

@article{Breakspear2017BrainDynamics,
  author  = {Breakspear, M.},
  title   = {Dynamic Models of Large-Scale Brain Activity},
  journal = {Nature Neuroscience},
  volume  = {20},
  number  = {3},
  year    = {2017}
}

@article{Deb2002NSGAII,
  author  = {Deb, K. and Pratap, A. and Agarwal, S. and Meyarivan, T.},
  title   = {A Fast and Elitist Multiobjective Genetic Algorithm: NSGA-II},
  journal = {IEEE Transactions on Evolutionary Computation},
  volume  = {6},
  number  = {2},
  year    = {2002}
}

@article{Fujii2017DisorderedQRC,
  author  = {Fujii, K. and Nakajima, K.},
  title   = {Harnessing Disordered-Ensemble Quantum Dynamics for Machine Learning},
  journal = {Physical Review Applied},
  volume  = {8},
  number  = {2},
  pages   = {024030},
  year    = {2017},
  doi     = {10.1103/PhysRevApplied.8.024030}
}

@article{GarciaBeni2023PhotonicQRC,
  author  = {Garcia-Beni, J. and Giorgi, G. L. and Soriano, M. C. and Zambrini, R.},
  title   = {Scalable Photonic Platform for Real-Time Quantum Reservoir Computing},
  journal = {Physical Review Applied},
  volume  = {20},
  pages   = {014051},
  year    = {2023},
  doi     = {10.1103/PhysRevApplied.20.014051}
}

@article{Honey2012SlowCortical,
  author  = {Honey, C. J. and Thesen, T. and Donner, T. H. and Silbert, L. J. and Carlson, C. E. and Devinsky, O. and Doyle, W. K. and Rubin, N. and Heeger, D. J. and Hasson, U.},
  title   = {Slow Cortical Dynamics and the Accumulation of Information over Long Timescales},
  journal = {Neuron},
  volume  = {76},
  number  = {2},
  pages   = {423--434},
  year    = {2012},
  doi     = {10.1016/j.neuron.2012.08.011}
}

@techreport{Jaeger2001EchoState,
  author = {Jaeger, H.},
  title  = {The ``Echo State'' Approach to Analysing and Training Recurrent Neural Networks},
  institution = {German National Research Center for Information Technology},
  year   = {2001},
  number = {GMD Report 148}
}

@article{Jaeger2004Harnessing,
  author  = {Jaeger, H. and Haas, H.},
  title   = {Harnessing Nonlinearity: Predicting Chaotic Systems and Saving Energy in Wireless Communication},
  journal = {Science},
  volume  = {304},
  pages   = {78--80},
  year    = {2004}
}

@article{Kutvonen2020OptimizingQRC,
  author  = {Kutvonen, A.},
  title   = {Optimizing a Quantum Reservoir Computer for Time Series Prediction},
  journal = {Scientific Reports},
  year    = {2020}
}

@article{Lachaux1999PhaseSynchrony,
  author  = {Lachaux, J.-P. and Rodriguez, E. and Martinerie, J. and Varela, F. J.},
  title   = {Measuring Phase Synchrony in Brain Signals},
  journal = {Human Brain Mapping},
  volume  = {8},
  number  = {1},
  pages   = {194--208},
  year    = {1999}
}

@article{Larger2017PhotonicRC,
  author  = {Larger, L. and Baylon-Fuentes, A. and Martinenghi, R. and Udaltsov, V. S. and Chembo, Y. K. and Jacquot, M.},
  title   = {High-Speed Photonic Reservoir Computing Using a Time-Delay-Based Architecture},
  journal = {Physical Review X},
  volume  = {7},
  pages   = {011015},
  year    = {2017},
  doi     = {10.1103/PhysRevX.7.011015}
}

@incollection{Lukosevicius2012PracticalESN,
  author    = {Lukosevicius, M.},
  title     = {A Practical Guide to Applying Echo State Networks},
  booktitle = {Neural Networks: Tricks of the Trade, Reloaded},
  editor    = {Montavon, G. and Orr, G. and Muller, K.},
  publisher = {Springer},
  year      = {2012}
}

@article{Lukosevicius2009RCReview,
  author  = {Lukosevicius, M. and Jaeger, H.},
  title   = {Reservoir Computing Approaches to Recurrent Neural Network Training},
  journal = {Computer Science Review},
  volume  = {3},
  number  = {1},
  pages   = {127--149},
  year    = {2009},
  doi     = {10.1016/j.cosrev.2009.03.005}
}

@article{Maass2002LiquidState,
  author  = {Maass, W. and Natschlager, T. and Markram, H.},
  title   = {Real-Time Computing Without Stable States},
  journal = {Neural Computation},
  volume  = {14},
  number  = {11},
  pages   = {2531--2560},
  year    = {2002}
}

@article{Nakajima2014SoftBodyRC,
  author  = {Nakajima, K. and Li, T. and Hauser, H. and Pfeifer, R.},
  title   = {Exploiting Short-Term Memory in Soft Body Dynamics as a Computational Resource},
  journal = {Interface},
  volume  = {11},
  pages   = {20140437},
  year    = {2014},
  doi     = {10.1098/rsif.2014.0437}
}

@article{Sakoe1978DTW,
  author  = {Sakoe, H. and Chiba, S.},
  title   = {Dynamic Programming Algorithm Optimization for Spoken Word Recognition},
  journal = {IEEE Transactions on Acoustics, Speech, and Signal Processing},
  volume  = {26},
  number  = {1},
  year    = {1978}
}

@misc{Sannia2025NonMarkovianQRC,
  author       = {Sannia, A. and Rodr{\'\i}guez, R. R. and Giorgi, G. L. and Zambrini, R.},
  title        = {Non-Markovianity and Memory Enhancement in Quantum Reservoir Computing},
  year         = {2025},
  eprint       = {2505.02491},
  archivePrefix= {arXiv},
  doi          = {10.48550/arXiv.2505.02491}
}

@article{Shenoy2021BrainWide,
  author  = {Shenoy, K. V. and Kao, J. C.},
  title   = {Measurement, Manipulation and Modeling of Brain-Wide Neural Population Dynamics},
  journal = {Nature Communications},
  volume  = {12},
  pages   = {633},
  year    = {2021},
  doi     = {10.1038/s41467-020-20371-1}
}

@article{Vlachas2020ForecastingRC,
  author  = {Vlachas, P. R. and Pathak, J. and Hunt, B. R. and Sapsis, T. P. and Girvan, M. and Ott, E. and Koumoutsakos, P.},
  title   = {Backpropagation Algorithms and Reservoir Computing in Recurrent Neural Networks},
  journal = {Neural Networks},
  volume  = {126},
  pages   = {191--217},
  year    = {2020},
  doi     = {10.1016/j.neunet.2020.02.016}
}

@article{Zhu2025FewAtomQRC,
  author  = {Zhu, C. and Ehlers, P. J. and Nurdin, H. I. and Soh, D.},
  title   = {Practical Few-Atom Quantum Reservoir Computing},
  journal = {Physical Review Research},
  volume  = {7},
  pages   = {023290},
  year    = {2025}
}

@article{ramezanian-panahi2022generative,
  title={Generative Models of Brain Dynamics},
  author={Ramezanian-Panahi, M. and Abrevaya, G. and Gagnon-Audet, J.-C. and Voleti, V. and Rish, I. and Dumas, G.},
  journal={Frontiers in Artificial Intelligence},
  volume={5},
  number={807406},
  year={2022},
  doi={10.3389/frai.2022.807406}
}

@article{manneschi2021parallelESN,
  title={Exploiting Multiple Timescales in Hierarchical Echo State Networks},
  author={Manneschi, L. and Ellis, M. O. A. and Gigante, G. and Lin, A. C. and Giudice, P. D. and Vasilaki, E.},
  journal={Frontiers in Applied Mathematics and Statistics},
  volume={6},
  number={616658},
  year={2021},
  doi={10.3389/fams.2020.616658}
}

@article{hamhoum2025multivariate,
  title={Multivariate Time Series Forecasting with Gate-Based Quantum Reservoir Computing on NISQ Hardware},
  author={Hamhoum, Wissal and Cherkaoui, Soumaya and Laprade, Jean-Fr{\'e}d{\'e}ric and Ahmed, Ola and Wang, Shengrui},
  journal={arXiv preprint arXiv:2510.13634},
  year={2025},
  primaryClass={cs.LG},
  url={https://arxiv.org/abs/2510.13634},
  doi={10.48550/arXiv.2510.13634}
}

@article{wolff2025quantum,
  title = {Quantum Computing for Neuroscience: Theory, Methods and Opportunities},
  author = {Wolff, A. and Choquette, A. and Northoff, G. and Iriki, A. and Dumas, G.},
  year = {2025},
  month = {September},
  day = {17},
  journal = {PsyArXiv},
  publisher = {OSF Preprints},
  doi = {10.31234/osf.io/vw8n3_v1},
  url = {https://doi.org/10.31234/osf.io/vw8n3_v1},
  note = {Preprint}
}

@article{bose_advancing_2026,
    title = {Advancing single-cell omics and cell-based therapeutics with quantum computing},
    volume = {27},
    copyright = {2026 Springer Nature Limited},
    issn = {1471-0080},
    url = {https://www.nature.com/articles/s41580-025-00918-0},
    doi = {10.1038/s41580-025-00918-0},
    language = {en},
    urldate = {2026-01-09},
    journal = {Nature Reviews Molecular Cell Biology},
    publisher = {Nature Publishing Group},
    author = {Bose, Aritra and Rhrissorrakrai, Kahn and Utro, Filippo and Parida, Laxmi},
    month = jan,
    year = {2026},
    pages = {394--408},
}

@article{flother_how_2024,
    title = {How quantum computing can enhance biomarker discovery for multi-factorial diseases},
    journal = {arXiv preprint arXiv:2411.10511},
    author = {Flöther, Frederik F and Blankenberg, Daniel and Demidik, Maria and Jansen, Karl and Krishnakumar, Rajiv and Laanait, Nouamane and Parida, Laxmi and Saab, Carl and Utro, Filippo},
    year = {2024},
    note = {Type: Journal Article},
}

@article{doga_how_2024,
    title = {How can quantum computing be applied in clinical trial design and optimization?},
    issn = {0165-6147},
    journal = {Trends in Pharmacological Sciences},
    author = {Doga, Hakan and Bose, Aritra and Sahin, M Emre and Bettencourt-Silva, Joao and Pham, Anh and Kim, Eunyoung and Andress, Alan and Saxena, Sudhir and Parida, Laxmi and Robertus, Jan Lukas},
    year = {2024},
    note = {Type: Journal Article},
}

@misc{javadiabhari2024quantumcomputingqiskit,
    author = {Ali Javadi-Abhari and Matthew Treinish and Kevin Krsulich and Christopher J. Wood and Jake Lishman and Julien Gacon and Simon Martiel and Paul D. Nation and Lev S. Bishop and Andrew W. Cross and Blake R. Johnson and Jay M. Gambetta},
    title = {Quantum computing with Qiskit},
      year={2024},
      eprint={2405.08810},
      archivePrefix={arXiv},
      primaryClass={quant-ph},
      url={https://arxiv.org/abs/2405.08810}, 
}

\vspace{12pt}
\color{red}

\end{document}